\documentclass[12pt]{amsart}

\usepackage{subfigure}
\usepackage{nicefrac}
\usepackage{latexsym,verbatim,amsbsy}
\usepackage[margin=1in]{geometry}  
\usepackage{amsmath}               
\usepackage{mathtools}             
\usepackage{amsfonts}              
\usepackage{amsthm}                
\usepackage{amssymb}               
\usepackage{dsfont}
\usepackage{bbm}
\usepackage{float}
\usepackage{mathrsfs}
\usepackage{url}
\usepackage{enumerate}

\usepackage[english]{babel}
\usepackage{times}

\usepackage{algorithm}
\usepackage{algorithmicx}
\usepackage{algpseudocode}

\usepackage{graphicx}              
\usepackage{wrapfig}
\usepackage{caption}
\usepackage{epstopdf}
\usepackage{tabularx}

\usepackage{enumerate}
\usepackage[mathlines]{lineno}

\newtheorem{theorem}{Theorem}[section]
\newtheorem{remark}[theorem]{Remark}
\newtheorem*{remark*}{Remark}

\newtheorem{lemma}[theorem]{Lemma}

\numberwithin{equation}{section}
\numberwithin{figure}{section}

\usepackage[dvipsnames]{xcolor}
\usepackage[colorlinks=true,
            linkcolor=red,
            urlcolor=magenta, 
            citecolor=red]{hyperref}
\definecolor{labelkey}{rgb}{0,0,1}

\newcommand{\mbf}[1]{\boldsymbol{#1}}
\newcommand{\lil}[1]{\ell_{#1}}
\newcommand{\innerprod}[2]{\langle #1, #2 \rangle}

\newcommand{\norm}[2]{|#1|_{#2}}

\newcommand{\realR}[1]{\mathbb{R}^{#1}}
\newcommand{\real}{\mathbb{R}}

\newcommand{\vunsub}{\mbf{x}}

\newcommand{\vtarget}{\vunsub_*}

\newcommand{\lone}{\vunsub_{BP}}
\newcommand{\ulm}[1]{\vunsub_{#1}}

\newcommand{\noise}{\mbf{\varepsilon}}

\newcommand{\nl}{\epsilon} 
\newcommand{\M}{m} 
\newcommand{\ba}{\mbf{a}}

\newcommand{\bu}{\mbf{u}}

\newcommand{\bv}{\mbf{v}}

\newcommand{\bx}{\mbf{x}}

\newcommand{\by}{\mbf{y}}
\newcommand{\bz}{\mbf{z}}

\newcommand{\X}{{W}} 
\newcommand{\w}{\mbf{w}}

\newcommand{\mK}{\mathcal{K}}

\newcommand{\mS}{\mathcal{S}}
\newcommand{\mT}{\mathcal{T}}

\newcommand{\result}{\by_*}
\newcommand{\nresult}{\result^{\nl}}

\newcommand{\resid}[1]{\mbf{r}_{\!#1}}

\let\emph\textit

\def\ks{k}  
\def\kk{k}   
\def\klam{s_\lambda} 
\newcommand{\sk}{\sigma_\ks} 
\newcommand{\slam}{s_{\lambda}} 

\newcommand{\betadelta}{{\beta_\delta}} 
\newcommand{\betaplustheta}{(\betadelta+\theta)} 
\newcommand{\deltap}{\delta} 
\newcommand{\betadeltap}{\beta_{\deltap}} 

\def\vdelta{0.7} 
\def\vdeltab{0.4}
\def\vtheta{0.1} 
\def\vbeta{2.42} 
\def\vchi{3.16} 
\def\vchisquared{11}  
\def\vchisquaredb{4}  
\def\vbetaplustheta{2.52} 
 
\def\vbetasqrtk{30.61} 
\def\vfracdelta{1.83}  
\def\vfracdeltabeta{4.43} 
\def\vthreefracdelta{5.48} 
\def\veta{0.59} 
\def\vlonebound{16.97}  
\def\veightbeta{24.04}  

\DeclareRobustCommand{\rchi}{{\mathpalette\irchi\relax}}

\newcommand{\irchi}[2]{\raisebox{\depth}{$#1\chi$}} 

\newcommand{\scr}{\kk_{\text{max}}}
\newcommand{\tik}[1]{\vunsub_{#1}} 

\newcommand{\sgn}[1]{\text{sgn}(#1)}
\newcommand{\vsgn}[1]{\,{\textbf{sgn}}(#1)}

\newcommand{\zero}{\mbf{0}}

\newcommand{\trans}[1]{#1^{\top}}

\newcommand{\argmin}[1]{\underset{#1}{\operatorname{arg}\operatorname{min}}\;}

\newcommand{\minprob}[2]{\argmin{#1}\Big\{#2\Big\}}
\newcommand{\compl}[1]{#1^c}

\newcommand{\supp}[1]{\text{supp}(#1)}

\newcommand{\mand}{\quad \text{and} \quad}

\newcommand{\ds}{\displaystyle}
\renewcommand{\geq}{\geqslant} 
\renewcommand{\leq}{\leqslant}  
\renewcommand{\ge}{\geqslant}
\renewcommand{\le}{\leqslant}

\def\LASSO{LASSO } 

\newcommand{\simon}[1]{{#1}}

\begin{document} 

\title[On the sparsity of \LASSO  minimizers]{{On the sparsity of \LASSO minimizers \\ in sparse   data recovery}}

\author{Simon Foucart}
\address{Department of Mathematics and Texas A\&M Institute of Data Science\newline \hspace*{0.4cm} Texas A\&M University, College Station, TX 77843}
\email{foucart@tamu.edu}

\author{Eitan Tadmor}
\address{Department of Mathematics and Institute for Physical Sciences \& Technology (IPST)\newline \hspace*{0.4cm} University of Maryland, College Park, MD 20742}
\email{tadmor@umd.edu}

\author{Ming Zhong}
\address{Texas A\&M Institute of Data Science\newline \hspace*{0.4cm} Texas A\&M University,
College Station, TX 77843}
\email{mingzhong@tamu.edu}

\date{\today}



\thanks{Research was supported in part by NSF grants CCF-1934904, DMS-2053172 (SF), and DMS16-13911 (ET) and ONR grants N00014-2012787 (SF) and N00014-2112773 (ET)}

\dedicatory{\bigskip{\large Dedicated to Ron DeVore with friendship and admiration}}

\begin{abstract}
We present a detailed analysis of the unconstrained $\ell_1$-weighted \LASSO 
method for  recovery of sparse data from its  observation  by randomly generated matrices, satisfying the Restricted Isometry Property (RIP) with constant $\delta<1$, and subject to negligible measurement and compressibility errors.
We prove that if the data is  $k$-sparse, then  the   size of support of the \LASSO minimizer, $s$, maintains  a comparable sparsity,  $s\leq C_\delta k$. For example, if $\delta=\vdelta$ then $s< \vchisquared k$ and a slightly smaller $\delta=\vdeltab$ yields $s< \vchisquaredb k$. We also derive  new $\ell_2/\ell_1$ error bounds which  highlight  precise dependence
  on $k$ and on the LASSO parameter $\lambda$,  before the error  is driven below the scale of  negligible  measurement/ and compressiblity errors.
\end{abstract}
\maketitle

\vspace*{-0.9cm}
\setcounter{tocdepth}{2}
\tableofcontents
\section{Introduction}
In 2006, the pioneering works  of Cand{\`e}s, Romberg and Tao \cite{CRT2006a,CRT2006b} and of Donoho \cite{donoho2006} suggested the framework of  a constrained $\lil{1}$-method to recover a sparse unknown $\vtarget \in \realR{N}$ from its  observation $\result = A\vtarget \in \realR{\M}$\footnote{Earlier  announcement of the works was presented in the workshops, organized together with Ron DeVore, which can be found at
\href{https://home.cscamm.umd.edu/programs/srs05/candes\_srs05.pdf}{home.cscamm.umd.edu/programs/srs05/candes\_srs05.pdf} and \href{https://home.cscamm.umd.edu/programs/srs05/donoho\_srs05.htm}{home.cscamm.umd.edu/programs/srs05/donoho\_srs05.htm}.}.  The key point is that one can design observing matrices $A \in \realR{\M\times N}$ with a relatively small number of observations,  $\M \ll N$, such that a constrained $\lil{1}$-method --- also known as Basis Pursuit (BP) in \cite{chen1994basis,chen1996basis,CDS1999} ---  finds a sparse solution as a minimizer  of\footnote{Given $x\in \realR{N}$ we let $\norm{\bx}{p}$ denote  its $\lil{p}$-norm, with the usual conventional limiting cases of $p=\infty$ and  $p=0$, where
$\displaystyle \norm{\bx}{\infty} := \max_{1 \le i \le N}|x_i|$,  and respectively $\norm{\bx}{0} := |\supp{\bx}|$ where $|\cdot|$ is the cardinality of a finite set.}
\[
  \lone \coloneqq \argmin{\bx \in \realR{N}}{\big\{\norm{\bx}{1} \ \big| \   A\bx = \result\}}, \qquad A\in \realR{m\times N}, \ m\ll N.
\]
This is closely related to the  well-known LASSO algorithm introduced in 1996   in  the  statistics literature \cite{tibshirani1996}, $\minprob{\norm{\bx}{1}\leq \delta}{\norm{\nresult-A\bx}{2}^2}$,  which  can be viewed as an $\ell_1$-penalty relaxation of a  least squares  subject to (possibly noisy) observation $\nresult$. 

The BP minimizer, $\lone$, recovers the sparse $\vtarget$ when the observing matrix $A$ satisfies an appropriate  recoverability condition; we mention here the Restricted Isometry Property (RIP) introduced in \cite{CRT2006a}, the $\lil{1}$-Coherence discussed in \cite{tropp2004, GN2003, DE2003, DH2001}, the restricted eigenvalue condition \cite[\S3]{bickel2009simultaneous}, or the Null Space Property (NSP)  of DeVore and his co-authors \cite{CDDV2009, DVPW2009}, and related Robust Null Sparse Property (RNSP) of \cite{FR2013}.  Important classes of such observing matrices with desired  sparse recoverability conditions are randomly generated, e.g., \cite[\S9]{FR2013}.
\subsection{Statement of main results}\label{sec:cs_tnp}
Throughout the paper we will be using the two notions of sparsity and compressibility. A vector $\bx \in \realR{N}$ is \emph{sparse} if 
\[
s_{\bx}:=\norm{\bx}{0} \ll N.
\]
In applications,  sparsity is often difficult to acquire, and clean observations are not always available, since the observation process is inevitably and easily corrupted by errors --- human and/or machine measurement errors.  We turn our attention to the recovery of {compressible} unknown from its \emph{noisy} observations.  
 A vector $\bx \in \realR{N}$ is \emph{compressible} of order $\kk$, or simply $\kk$-compressible, if its content is faithfully captured by a $\kk$-sparse vector --- specifically, if   its  $\lil{1}$-distance  to  the set of all $\kk$-sparse vectors,
  \begin{equation}\label{eq:bestksparse}
    \sigma_\kk(\bx) := \inf_{\bz \in \realR{N}}\left\{\norm{\bx- \bz}{1} :  \norm{\bz}{0} \le k\right\},
  \end{equation}
  is small relative to  $\norm{\bx}{1}$.
  We note that $\sigma_\kk(\bx)$ is realized by a (not necessarily unique) vector, denoted $\bx(\kk)$, whose non-zero entries are the $\kk$ largest  of $\bx$ in absolute value. 

\smallskip\noindent
Let $\vtarget$ be a compressible unknown of order $\ks$ so that $\sigma_\ks(\vtarget) \ll \norm{\vtarget}{1}$, and assume we only have access  to its measured observation $\nresult = A\vtarget + \noise$.  The  term $\noise$ is the measurement error caused by a number of factors which are assumed statistically independent of the unknown $\vtarget$ and the observing operator  $A$. The details of $\noise$ remain untraceable except for its size  which is  assumed to
be negligibly small. 
In this case, one should not expect  an exact recovery of a sparse $\vtarget$, but instead,  accept an approximate solution, $\nresult = A\vtarget(\ks) + \noise'$, where $\noise'=A(\vtarget-\vtarget(\ks))+\noise$ is adapted to the  small scale built into the problem, which consists of two contributions --- 
the small measurement error, $\nl:=\norm{\noise}{2}$,  
and the small compressibility error,\footnote{The columns of $A$ are assumed $\ell_2$-normalized so that $\norm{A}{1\rightarrow2}=1$.} $\norm{A(\vtarget-\vtarget(\ks))}{2}\leq \sigma_\ks(\vtarget)$,
\[
\nresult = A\vtarget(\ks) + \noise', \qquad \norm{\noise'}{2}\leq \mu,  
\]  
such that  $\mu=\sigma_\ks(\vtarget)+\nl$ is much smaller relative to the unknown data, $\mu \ll \norm{\vtarget}{1}$.
\smallskip\noindent
  Although the observing operator $A$ is linear, the recovery of $\vtarget$ by a direct ``solution'' of the linear problem $A\bx = \nresult$ is ill-posed, unless  additional conditions on  $A$ and $\vtarget$ are enforced so that the unknown object $\vtarget$, or at least a faithful approximation of it, is recovered by solving an augmented  \emph{well-posed} regularized minimization problem. On the way, the original linear problem is replaced by a nonlinear procedure.
To capture the compressible information of $\vtarget$ from its noisy observation
$\nresult$, we \simon{seek}  a  minimizer of the unconstrained $\lil{1}$-regularized Least Squares problem,
\begin{equation}\label{eq:cs_uncon_l1}
  \ulm{\lambda} \coloneqq \minprob{\bx \in \realR{N}}{\lambda\norm{\bx}{1} + \frac{1}{2}\norm{\nresult - A\bx}{2}^2}, \qquad A\in \realR{m\times N}, \ m\ll N.
\end{equation}
The unconstrained variational statement  \eqref{eq:cs_uncon_l1}  falls under the general class of  Tikhonov regularization. The distinctive feature is the $\ell_1$-regularization, leading to an approximate decomposition of the basis pursuit of Chen \& Donoho   \cite{chen1994basis},
$\nresult=A\ulm{\lambda} +\resid{\lambda}$ with (hopefully) small
residual, $\resid{\lambda}= \nresult - A\ulm{\lambda}$,
depending on a parameter $\lambda$. This  version of  $\ell_1$-regularization, called ``Basis Pursuit De-Noising'' in \cite{chen1996basis},  which became known as the unconstrained $\ell_1$-weighted LASSO, is the main focus of our work. 
As noted in the 1996 thesis \cite{chen1996basis}, the work on this version of BP  was motivated by a series of ideas using $\ell_0/\ell_1$-based regularization that appeared in early 1990s, primarily the empirical atomic decomposition of Donoho and Johnstone \cite{donoho1995empirical},  $\minprob{\bx \in \realR{N}}{\lambda\norm{\bx}{0} + \frac{1}{2}\norm{\nresult - A\bx}{2}^2}$, the multi-scale edge representation  with wavelets of Hwang \simon{and} Mallat \cite{mallat1992singularity} and  the TV-based de-noising method of ROF \cite{rudin1992nonlinear}, 
$\minprob{\bx \in \realR{N}}{\lambda\norm{\bx}{TV} + \frac{1}{2}\norm{\nresult - A\bx}{2}^2}$. These works  were later further explored  as the Lagrangian formulation of the quadratically constrained Basis Pursuit de-noising \cite{CDS1999,CRT2006b} and the noise-aware $\lil{1}$-minimization \cite{FR2013}.

 Since  $\lambda>0$ controls the distance between $A\tik{\lambda}$ and $\result$, the parameter $\lambda$ can be interpreted as a regularization \emph{scale}.  In a subsequent work, \cite{TZII2021}, we pursue  a \emph{multi-scale} generalization based on a ladder of hierarchical scales constructed by the  Hierarchical Decomposition (HD) method   \cite{TNV2004, TNV2008,TT2010, Tadmor2015}.   The goal of this work is to analyze the sparsity behavior of the \emph{mono-scale} \LASSO  \eqref{eq:cs_uncon_l1}, 
  observed by a sub-class of RIP matrices satisfying the Robust Null Space Property (RNSP)   which is discussed in section \ref{sec:robust}.
Our main results,  outlined and proved in section \ref{sec:main}, are summarized in the following. Our results involve three main parameters: the Restricted Isometry Constant (RIC) in \eqref{eq:RIP} below, $\delta=\delta_{k}<1$,  the related RNSP constant, $\ds \betadelta=\frac{\sqrt{1+\delta}}{\sqrt{1-\delta^2}-\nicefrac{\delta}{4}}$, depending on the RIC $\delta$, and the small scale of compressibility$+$measurement, $\ds \mu=\sigma_\ks(\vtarget)+\nl$.
\begin{theorem}[{\bf Main result}]\label{thm:Main} 
Let $\vtarget$ be $\kk$-compressible, and let $\nresult = A\vtarget + \noise$ be its  observation  with observing matrix $A$ satisfying the RIP \simon{\eqref{eq:RIP}}  with constant $\delta$ large enough, $\delta>\delta_t$, such that \eqref{eq:whatist} below holds.  
Let $\ulm{\lambda}$ be the \LASSO minimizer \eqref{eq:cs_uncon_l1}.

\begin{enumerate}[\mbox{      } (i)]
 \item{\bf (Sparsity)}. The sparsity of the \LASSO minimizer,  
$\slam=s_{\ulm{\lambda}}$, does not exceed
  \[
  \slam <(1+\delta) \Big(\betadelta\sqrt{k}+\frac{2\mu}{\lambda}\Big)^2.
  \]
   \end{enumerate}
 \begin{enumerate}[\!\!(ii)] 
   \item{\bf ($\ell_2$-error bound)}. The following $\ell_2$-error bound holds 
 \[
\frac{1}{\sqrt{1+\delta}}\Big(\frac{\sqrt{\slam}\lambda}{\sqrt{1+\delta}} -
\mu\Big)\leq  \norm{\ulm{\lambda}-\vtarget(\ks)}{2} \leq \frac{1}{\sqrt{1-\deltap}}\big(\betadeltap\sqrt{k}\lambda+3\mu\big).
\]
   \end{enumerate}
 \begin{enumerate}[\hspace*{-0.4cm}(iii)]
 \item{\bf ($\ell_1$-error bound)}. The following $\ell_1$-error bound holds
 \[
 \norm{\ulm{\lambda}-\vtarget(\ks)}{1} < \frac{\sqrt{1+\delta}}{\sqrt{1-\delta}}\frac{1}{\lambda}\big((\betadelta + \nicefrac{1}{2})\sqrt{k}\lambda + 2\mu\big)^2.
\]
   \end{enumerate}
    We interpret these bounds as follows. Set $\theta =\nicefrac{2\mu}{\sqrt{k}\lambda}$, then (i) reads
  \[
  \slam \leq \rchi^2 k, \qquad \rchi=\sqrt{1+\delta}(\betadelta+\theta).
\]
 Thus, if  $\theta\leq 1$ --- namely, as long as $\lambda$ does not get exceedingly small so that  $\lambda \geq \nicefrac{2\mu}{\sqrt{k}}$, then the sparsity of $\ulm{\lambda}$ is comparable to the sparsity of $\vtarget$. Furthermore, in (ii) we have the $\ell_2$-error bound of order $\lesssim \sqrt{k}\lambda+\mu$ and in (iii), an $\ell_1$-error bound of order   $\lesssim k\lambda+ \sqrt{k}\mu$.
\end{theorem}
  We conclude with a few comments on theorem \ref{thm:Main}. The sparsity bound in (\emph{i}) with RIC  $\delta=\vdelta$ yields $\slam < \vchisquared k$, while a slightly smaller RIC $\delta=\vdeltab$ yields $\slam < \vchisquaredb k$. 
This should be compared with the sparsity bounds in  \cite[Theorem 3]{belloni2013least} and \cite{su2017false}.
  The $\ell_2$-upper bound on the right of (\emph{ii})  is not new; here we  recover the $\ell_2$-bound, derived  under appropriate assumptions, in \cite{candes2007dantzig}, \cite[Theorem 7.1]{bickel2009simultaneous} and
\cite{meinshausen2009lasso,tang2011performance,hastie2015statistical}.  This should be contrasted with the  $\ell_2$-error \emph{lower-bound} on the   left, derived in section \ref{sec:elltwo}  (see figure \ref{fig:simulEnt}). 
Indeed, this $\ell_2$ lower-bound is the essential ingredient in our proof of the  sparsity bound in ($i$).
Finally, the $\ell_1$-bound in (\emph{iii}) with  RIC  $\delta=\vdelta$ yields  $\norm{\ulm{\lambda}-\vtarget(\ks)}{1}<\vlonebound k\lambda + \veightbeta\sqrt{k}\mu$.
   Here, the linear decay with $\lambda$ is not new and can be found for example, under various assumptions,  in \cite[Theorem 7.1]{bickel2009simultaneous} and \cite[Theorem 6.1]{buhlmann2011statistics}. 

\section{The Robust Null Space Property}\label{sec:robust}

\noindent
{\bf Optimality of the minimizer}. The variational problem \eqref{eq:cs_uncon_l1} admits a minimizer, $\ulm{\lambda}$, and at least for certain relevant classes of full row rank $A$'s, the minimizer is unique, \cite{ZYC2015}. The minimizer is completely characterized by its residual, $\resid{\lambda}:= \nresult - A\ulm{\lambda}$ (to simplify notations we suppress the dependence of $\resid{\lambda}$ on $\nl$). We summarize the results from  \cite[\S2.1]{TNV2008},\cite[Appendix]{Tadmor2015} where we distinguish between two cases.
  
\renewcommand{\labelenumi}{({\bf \roman{enumi}})}
 \begin{enumerate}
       \item\label{ietm:i} If $\lambda \ge \lambda_\infty:=\norm{\trans{A}\nresult}{\infty}$ then  \eqref{eq:cs_uncon_l1} admits only the trivial minimizer $\ulm{\lambda} \equiv \zero$.
 In this case, $\lambda$ is too large to extract the compressibility information in $\nresult$.
 
\item\label{item:ii} If $\lambda < \lambda_\infty=\norm{\trans{A}\nresult}{\infty}$ then \eqref{eq:cs_uncon_l1} admits a non-trivial  minimizer, $\ulm{\lambda}$, with the corresponding  residual, $\resid{\lambda}= \nresult - A\ulm{\lambda}$, such that $(\ulm{\lambda},\resid{\lambda})$ forms an \emph{extremal pair} in the sense that
\begin{equation}\label{eq:resideq}
 \innerprod{A\ulm{\lambda}}{\resid{\lambda}} = \lambda\norm{\ulm{\lambda}}{1} \ \ \text{and} \ \ \norm{\trans{A}\resid{\lambda}}{\infty}=\lambda.
  \end{equation}
  \end{enumerate}
  To proceed we will need the following notations.
   The restriction of a vector $\w \in \realR{N}$ on an index set $\mK \subset \{1,2,\ldots,N\}$ of size $k=|\mK|$ is denoted  ${\w}_{\mK}:=\{w_i, \  i \in \mK\} \in \realR{k}$.
 Similarly, given a matrix $\X \in \realR{\M\times N}$ with columns  $\w_1,\w_2,\ldots$, its restriction on an index set $\mK$ of size $k=|\mK|$ consists of the $k$ columns $\displaystyle \X_{\mK} := \text{col}\{\w_i,  i \in \mK\}$.
 The size of $\X$ can be measured by its induced  matrix norm, $\displaystyle \|\X\|_{p}=\mathop{\sup}_{\norm{\w}{p} = 1}\norm{\X\w}{p}$.
 The signum vector is defined component-wise,  $\vsgn{\w}_i = \sgn{w_i}$, in terms of the usual  signum function 
$\displaystyle   \sgn{w} = \left\{
     \begin{array}{rr}
       -1, &  w < 0 \\
       1, &  w > 0
      \end{array}
   \right\}$ for $w\neq 0$.

\smallskip\noindent
{\bf Restricted Isometry Poperty (RIP)}.
A matrix $A$ satisfies the Restricted Isometry Property (RIP) of order $k$ with Restricted Isometry Constant (RIC) $\delta_k<1$ if the following holds,  \cite{CT2007,donoho2006,CRT2006a, BCT2011},
\begin{equation}\label{eq:RIP}
(1-\delta_k)|\bx|^2_2 \leq |A\bx|^2_2 \leq (1+\delta_k)|\bx|^2_2, \qquad \forall\norm{\bx}{0}\leq k.
\end{equation}
Throughout the paper we adopt the usual assumption that $\delta_k$ is measured for $A$'s with  $\ell^2$-normalized \emph{columns}\footnote{The RIP of $A$ asserts that for any subset  of its $k$ columns, $\{\ba_i\}_{i\in\mK}$, the  entries $|\langle \ba_i,\ba_j\rangle|_{i\neq j}\lesssim \delta_k$ while $|\ba_i|^2_2=1+\epsilon_i$ such that $|\epsilon_i|\lesssim \delta_k$. Therefore, one can always  re-normalize the columns of $A$ by a factor$\lesssim (1-\delta_k)^{-1/2}$ yielding a new  RIP matrix with $\ell_2$-normalized columns and with possibly slightly larger  RIP constant $\delta'_k \lesssim \delta_k/(1-\delta_k)$.}.
There are two classes of matrices $A \in \real^{\M \times N}$ satisfying the RIP of order $k$: deterministic  $A$'s with number of observations $\M \gtrsim k^2$ (the quadratic bottleneck is lessened in \cite{BDFKK2011}); and a large class of randomly generated $A$'s for which the restriction on the number of observations can be further lessened to  having only $\M$ observations, \cite[\S9.4]{FR2013}
\begin{equation}\label{eq:whatisdelta}
\M \sim Const\cdot\delta^{-2}k\ln\big(\nicefrac{eN}{k}\big).
\end{equation}
Cand\`{e}s proved the exactness of the constrained BP for RIP matrices  with $\delta <\sqrt{2}-1$, \cite{candes2008restricted}. 
Further refinements \simon{were reported in \cite{foucart2012sparse} before the definitve result of \cite{cai2013sparse}}.

\smallskip\noindent
{\bf \simon{Robust Null Space Property} (RNSP)}. 
A crucial step in quantifying the recovery error of $\vtarget$ using \eqref{eq:cs_uncon_l1} is to enforce a recoverability condition on the observing matrix $A$. This brings us to the  Robust Null Sparse Property (RNSP) introduced in \cite[\S4.3]{FR2013}.  
A matrix $A \in \realR{\M\times N}$ satisfies the RNSP of order $k$ with constants $0 < \rho < 1$ and $\tau > 0$, if for all  $\mK \subset \{1,2,\ldots, N\}$ of size $|\mK|\leq k$, there holds 
\begin{equation}\label{eq:RNSPk}
\norm{\bx_{\mK}}{1} \le \rho \norm{\bx_{\compl{\mK}}}{1} + \tau\norm{A\bx}{2}, \qquad \forall \bx \in \realR{N}.
\end{equation}
We refer to the   ``RNSP${}_{\rho,\tau}$ of order $k$'', and unless needed, we suppress the dependence of ($\rho,\tau)$ on $k$. In particular, given a $k$-sparse $\bv$ and any $\bu$, we apply \eqref{eq:RNSPk} to $\bx=\bu-\bv$ with 
$\mK=\supp{\bv}$, where  $\norm{\bx_{\mK}}{1}- \norm{\bx_{\compl{\mK}}}{1}\geq \norm{\bv}{1} - \norm{\bu}{1}$ yields the following useful consequence of RNSP.
\begin{lemma}
 If $A\in \realR{\M\times N}$ satisfies the RNSP${}_{\rho,\tau}$ of order $k$, then for all $k$-sparse $\bv$'s and any $\bu$,
\begin{equation}\label{eq:rearrange}
\norm{\bv}{1} - \norm{\bu}{1} \leq \tau\norm{A(\bu-\bv)}{2}, \qquad |\supp{\bv}|\leq k.
\end{equation}
\end{lemma}
As an example for the class of observation \simon{matrices} satisfying the RNSP${}_{\rho,\tau}$ of order $k$,  we  mention the  class  of randomly generated RIP matrices  with RICs $ \delta=\delta_{2k}$, \cite[Theorem 6.13]{FR2013}, 
\begin{equation}\label{eq:RNSP}
\rho = \frac{\delta}{\sqrt{1 - \delta^2} - \nicefrac{\delta}{4}} \mand \tau= \beta\sqrt{k}, \quad \beta:= \frac{\sqrt{1 + \delta}}{\sqrt{1 - \delta^2} - \nicefrac{\delta}{4}}, \qquad \delta=\delta_{2k}.
\end{equation}
These RNSP parameters,  $(\rho, \beta)$, are dictated as increasing functions of the RIC $\delta< 1$.  A smaller $\delta$ requires an increased number of observations. 
All proofs invoke different classes of observing matrices which  are randomly generated so that they satisfy a desirable observing  properties--- RIP, RNSP, or Constrained Minimal Singular Values (CMSV) property. Accordingly, the error statements are probabilistic in nature, referring to the ensemble of these observations. 

\section{On the sparsity of the unconstrained LASSO minimizer}\label{sec:main}
We analyze the sparsity and $\ell_1/\ell_2$-error bounds of the minimizer \eqref{eq:cs_uncon_l1} in recovering $\vtarget(\ks)$ from the  observation $\nresult=A\vtarget+\noise$, with small measurement error, $\nl=\norm{\noise}{2}$, and --- assuming that $\vtarget$ is $\ks$-compressible --- with small  compressibility error, $\sigma_k(\vtarget) =\norm{\vtarget-\vtarget(\ks)}{1}$. 
Set
\[
 \mu:=\sk(\vtarget)+\nl.
\]
 Clearly, since the exact solution  is observed up to $\ell_2$ residual error of order $\norm{\nresult-A\vtarget}{2}\leq \mu$, we do not have much to say when the computed residual error $\norm{\resid{\lambda}}{2}$ is of order  $\mu $ and we   will therefore limit ourselves to the parametric regime where $ \norm{\resid{\lambda}}{2}\gg \mu$. Below we show that $\norm{\resid{\lambda}}{2}\sim \lambda\sqrt{k}$ and therefore throughout the paper we make  the assumption 
\begin{equation}\label{eq:mu}
 \theta :=\frac{2\mu}{\lambda\sqrt{k}}  \leq 1, \qquad \mu=\sk(\vtarget)+\nl.
 \end{equation}
Thus, we assume  the LASSO weight, $\lambda$, does not get exceedingly small,  $\lambda\geq \nicefrac{2\mu}{\sqrt{k}}$. In concrete examples demonstrating the  sparsity and error bounds reported below we use $\theta= \vtheta$, corresponding to  $\lambda\geq \nicefrac{20\mu}{\sqrt{k}}$.
\begin{lemma}[{\bf The re-scaled residual --- an upper-bound}]\label{lem:res2}
Fix  $\lambda < \lambda_\infty:=|\trans{A}\nresult|_\infty$. Let $\nresult = A\vtarget + \noise$ be the  observation of a $\ks$-compressible unknown $\vtarget \in \realR{N}$, observed by $A \in \realR{\M\times N}$ satisfying the RNSP${}_{\rho,\tau}$ of order $\ks$, \eqref{eq:RNSP}. Let $\mu$ denote the small scale of  $\ks$-compressiblity  and measurement errors, \simon{see \eqref{eq:mu}}. Then the residual of the \LASSO \eqref{eq:cs_uncon_l1}, $\resid{\lambda}=\nresult-A\ulm{\lambda}$, satisfies
  \begin{equation}\label{eq:res2error}
\frac{\norm{\resid{\lambda}}{2}}{\lambda} \leq \betaplustheta\sqrt{k},
\qquad \betadelta= \frac{\sqrt{1 + \delta}}{\sqrt{1 - \delta^2} - \nicefrac{\delta}{4}}.
\end{equation}
\end{lemma}
\noindent

\begin{proof} 
Clearly, $\norm{A(\ulm{\lambda} -\vtarget(\ks))}{2}\leq \norm{\resid{\lambda}}{2}+\mu$.
Using \eqref{eq:rearrange} with the $k$-sparse $\bv=\vtarget(\ks)$ and $\bu=\ulm{\lambda}$ yields
 \begin{equation}\label{eq:xxr}
 \norm{\vtarget(\ks)}{1}  - \norm{\ulm{\lambda}}{1} \leq \tau \norm{A(\ulm{\lambda} -\vtarget(\ks))}{2}\leq \tau\norm{\resid{\lambda}}{2}+\tau\mu.
 \end{equation}
 Next, a lower-bound for the quantity on the left follows. Recall that   $\ulm{\lambda}$, being the \LASSO minimizer \eqref{eq:cs_uncon_l1}, is characterized  by the extremal property that its scaled residual $\ds \bz=\frac{\resid{\lambda}}{\lambda}$ satisfies \eqref{eq:resideq}, 
 \[
\norm{\ulm{\lambda}}{1} = \langle A\ulm{\lambda},\bz\rangle \ \ \text{and} \ \ \norm{\trans{A}\bz}{\infty}=1, \qquad \bz:=\frac{\resid{\lambda}}{\lambda}.
 \]
 Hence 
 \[
 \begin{split}
 \norm{\vtarget(\ks)}{1}  - \norm{\ulm{\lambda}}{1}
 & \geq \langle \vtarget(\ks),\trans{A}\bz\rangle - \langle A\ulm{\lambda},\bz\rangle = \langle A \vtarget(\ks)- A\ulm{\lambda},\bz\rangle \\
  & = \langle\resid{\lambda},\bz\rangle + \langle A\vtarget(\ks)-\nresult,\bz\rangle   \\
  &   \geq \frac{\norm{\resid{\lambda}}{2}^2}{\lambda} -\norm{A\vtarget(\ks)-\nresult}{2}\frac{\norm{\resid{\lambda}}{2}}{\lambda}.
 \end{split}
 \]
Now,  assumption  \eqref{eq:mu} and the fact that $\norm{A}{1\rightarrow 2}\leq 1$ imply,
\[
\norm{A\vtarget(\ks)-\nresult}{2}\leq \norm{A\vtarget-\nresult}{2} + \norm{A(\vtarget(\ks)-\vtarget)}{2} \leq \nl +\sk(\vtarget)=\mu,
\]  
and we end with the desired lower-bound
\begin{equation}\label{eq:xxl}
\norm{\vtarget(\ks)}{1}  - \norm{\ulm{\lambda}}{1}\geq
\frac{\norm{\resid{\lambda}}{2}^2}{\lambda}-\mu\frac{\norm{\resid{\lambda}}{2}}{\lambda}.
\end{equation}
Combining \eqref{eq:xxr} and \eqref{eq:xxl} we conclude that $\ds \norm{\bz}{2}=\frac{\norm{\resid{\lambda}}{2}}{\lambda}$ satisfies the quadratic inequality,
$\ds \norm{\bz}{2}^2\leq \Big(\tau+\frac{\mu}{\lambda}\Big)\norm{\bz}{2} +\tau \frac{\mu}{\lambda}$, and therefore
\begin{equation}\label{eq:halftheta}
\frac{\norm{\resid{\lambda}}{2}}{\lambda}=\norm{\bz}{2} <  \tau+\frac{2\mu}{\lambda}= \betaplustheta\sqrt{k},
\end{equation}
proving \eqref{eq:res2error}.
\end{proof}

\subsection{Bounds of the sparsity}\label{sec:sparsity} 
 We now come to the main point of the  lower-bound    on the scaled residual in terms of the size of the support of $\ulm{\lambda}$, $\displaystyle \frac{\norm{\resid{\lambda}}{2}}{\lambda}\gtrsim \sqrt{\klam}$. Fix  $\lambda < \lambda_\infty:=|\trans{A}\nresult|_\infty$. Recall that if   $\ulm{\lambda}$ is the \LASSO minimizer \eqref{eq:cs_uncon_l1}  then by the extremal property \eqref{eq:resideq}, the scaled residual   $\displaystyle \bz=\frac{\resid{\lambda}}{\lambda}$ satisfies the two properties $\langle A\ulm{\lambda},\bz\rangle=\norm{\ulm{\lambda}}{1}$  and $\norm{\trans{A}\bz}{\infty}=1$. Thus, the extremal   $\ulm{\lambda}$ with support $\mS=\supp{\ulm{\lambda}}$ of size $\klam=\norm{\ulm{\lambda}}{0}$, is identified by a re-scaled residual satisfying 
 \begin{equation}\label{eq:B}
 (\trans{A}\bz)_\mS= \vsgn{\ulm{\lambda,\mS}}, \qquad \bz=\frac{\resid{\lambda}}{\lambda}, \quad \mS=\supp{\ulm{\lambda}}.
 \end{equation}
 Fix the integer $t$,
 \begin{equation}\label{eq:whatist}
 t:=[(1+\delta)\betaplustheta^2 k]+1 \ \ \textnormal{with constant} \ \ \delta > \delta_{t}.
 \end{equation}
   Since the RIC $\delta_{t}$ is increasing with the order $t$, there is no need to trace a precise fixed point  associated with  \eqref{eq:whatist}, $t=[(1+\delta_t)(\beta_{\delta_t}+\theta)^2 k]+1$. Instead, we can use a priori bounds of $\delta_t$; for example, if we restrict ourselves to the range $\delta<\vdelta$, we can set  the integer upper bound $t=\vchisquared k $. Below, we demonstrate  refined versions of this bound.\newline
 We claim that 
 \begin{equation}\label{eq:sltk}
 \klam < t=[(1+\delta)\betaplustheta^2 k]+1.
 \end{equation}
 To this end we proceed by contradiction. Assume $\klam \geq t$. Then  the support of $\ulm{\lambda}$ has a subset $\mT$ of size $t$ for which 
 the extremal property \eqref{eq:B} reads $(\trans{A}\bz)_\mT= \vsgn{\ulm{\lambda,\mT}}$, and the RIP \eqref{eq:RIP} for such set $\mT$  implies
\begin{equation}\label{eq:upperAAT}
\begin{split}
\norm{A(\trans{A}\bz)_\mT}{2}^2   
 \leq (1+\delta_t)\norm{(\trans{A}\bz)_\mT}{2}^2 = (1+\delta_t)\norm{\vsgn{\ulm{\lambda,\mT}}}{2}^2=(1+\delta_t)t.
 \end{split}
\end{equation}
On the other hand, we have
\[
\norm{A(\trans{A}\bz)_\mT}{2}^2 \geq \frac{1}{\norm{\bz}{2}^2}\big\langle A(\trans{A}\bz)_\mT,\bz\big\rangle^2 =\frac{1}{\norm{\bz}{2}^2}\big\langle (\trans{A}\bz)_\mT, \trans{A}\bz\big\rangle^2= \frac{1}{\norm{\bz}{2}^2}
\norm{(\trans{A}\bz)_\mT}{2}^4 = \frac{t^2}{\norm{\bz}{2}^2}.
\]
The last two inequalities  followed by Lemma \ref{lem:res2} imply 
$t\leq (1+\delta_t)\norm{\bz}{2}^2 < (1+\delta)\betaplustheta^2k$,
which contradicts the definition of $t$,
\[
t=[(1+\delta)\betaplustheta^2 k]+1 \geq (1+\delta_t)\betaplustheta^2k.
\]
 Thus, \eqref{eq:sltk} holds.\newline
 In fact, a refined  statement follows. Now that we know $|\mS|\leq[(1+\delta)\betaplustheta^2k]$ we can argue along the same line as above with $\mT=\mS$, obtaining $\klam\leq (1+\delta)\norm{\bz}{2}^2$.  
 \begin{lemma}[{\bf The re-scaled residual --- a lower-bound}]\label{lem:res3}
 Fix  $\lambda < \lambda_\infty:=|\trans{A}\nresult|_\infty$ and let  $\ulm{\lambda}$ be the  $\slam$-sparse minimizer of the corresponding \LASSO \eqref{eq:cs_uncon_l1}, observed with RIC $\delta$ such that \eqref{eq:whatist} holds. Then the residual, $\resid{\lambda}=\nresult-A\ulm{\lambda}$, satisfies
\begin{equation}\label{eq:lower}
\frac{\norm{\resid{\lambda}}{2}^2}{\lambda^2} \geq \frac{\klam}{1+\delta}.
\end{equation} 
\end{lemma}
\noindent
Combining the lower- and upper-bounds of $\displaystyle  \frac{\norm{\resid{\lambda}}{2}}{\lambda}$ we conclude the following.
\begin{theorem}[{\bf Sparsity bound}]\label{thm:main}
Fix $\lambda < \lambda_\infty:=|\trans{A}\nresult|_\infty$ and let  $\ulm{\lambda}$ be the  $\slam$-sparse minimizer of the corresponding \LASSO \eqref{eq:cs_uncon_l1}, observed with RIC $\delta$ such that \eqref{eq:whatist} holds.
Then
\begin{equation}\label{eq:sparsity}
\frac{\slam}{1+\delta} \leq \frac{\norm{\resid{\lambda}}{2}^2}{\lambda^2} \leq \betaplustheta^2k, \qquad \delta>\delta_t, \quad t=[(1+\delta_t)(\beta_{\delta_t}+\theta)^2 k]+1.
\end{equation}
In particular, we recover  \eqref{eq:sltk}, $\slam\leq [(1+\delta)\betaplustheta^2 k]$.
\end{theorem}

We demonstrate the application of corollary \ref{thm:main} for different choices of RICs. In all cases, we set  $\theta=\vtheta$. We begin with  the RIC  $\delta=\vdelta$, obtaining $\betaplustheta=\vbetaplustheta \leadsto 
\klam\leq (1+\delta)\betaplustheta^2 k  <\vchisquared k$. Thus, with $t=\vchisquared k$ we require $\delta_{\vchisquared k}<\vdelta $ which in turn, by \eqref{eq:whatisdelta}, set the number of required observations 
\[
\klam < \vchisquared k: \qquad  \M \approx Const. \frac{\vchisquared k}{\vdelta^2} \ln(\nicefrac{eN}{k}) \approx  Const. \simon{22.4}\,k\ln(\nicefrac{eN}{k}). 
\]  
For a second example we choose a smaller RIC $\delta=\vdeltab$ and  $\theta=\vtheta$. Recall, that a smaller $\delta$ requires more observations yet in the number of observations in the present context depends on $\delta_t$. In this case $\betaplustheta=1.55 \leadsto 
\klam\leq (1+\delta)\betaplustheta^2 k =3.36k  <\vchisquaredb k$. This requires  a slightly larger number of observations (or at least a smaller bound  \eqref{eq:whatisdelta})
\[
\klam<\vchisquaredb k : \qquad \M \approx Const. \frac{\vchisquaredb k}{\vdeltab^2} \ln(\nicefrac{eN}{k}) \approx Const. 25\,k \ln(\nicefrac{eN}{k}).
\]
Finally, as a third example we choose an even smaller the RIC $\delta=0.26$ and the same  $\theta=\vtheta$.  In this case $\betaplustheta=1.35 \leadsto 
\klam\leq (1+\delta)\betaplustheta^2 k =2.28k  <3k$, and this yields  the number of required observations 
\[
\klam<3k : \qquad \M \approx Const. \frac{3k}{0.26^2} \ln(\nicefrac{eN}{k}) \approx Const. 44.4\, k\ln(\nicefrac{eN}{k}).
\]

\begin{remark}[{\bf On the threshold parameter $\rchi$}]\label{rem:rchi} 
Observe that the sparsity bound $\klam$ is uniform in the small scale  $\mu$ throughout the parametric regime assumed in \eqref{eq:mu}. Thus, in the range of $\lambda \gg \nicefrac{2\mu}{\sqrt{k}}$, the support of the computed solution $\norm{\ulm{\lambda}}{0}$ can grow  at most by a  fixed factor  relative to the $k$-support of underlying unknown $\vtarget$, 
\cite[Appendix A]{tang2011performance}. We write  
\begin{equation}\label{eq:smax}
\slam<([\rchi^2]+1) k, \qquad \rchi:= \sqrt{1+\delta}\betaplustheta=\frac{1+\delta}{\sqrt{1-\delta^2}-\nicefrac{\delta}{4}}+\sqrt{1+\delta}\theta.
\end{equation}
We have the theoretical bounds $[\rchi^2]+1  = \vchisquared$ corresponding to $\delta=\vdelta$ and   $[\rchi^2]+1 = 4$ corresponding to $\delta\approx \vdeltab$. 
\end{remark}

\subsection{Numerical simulations}

\begin{figure}[H]
  \centering
  \includegraphics[width=14.8cm,height=12.8cm]
  {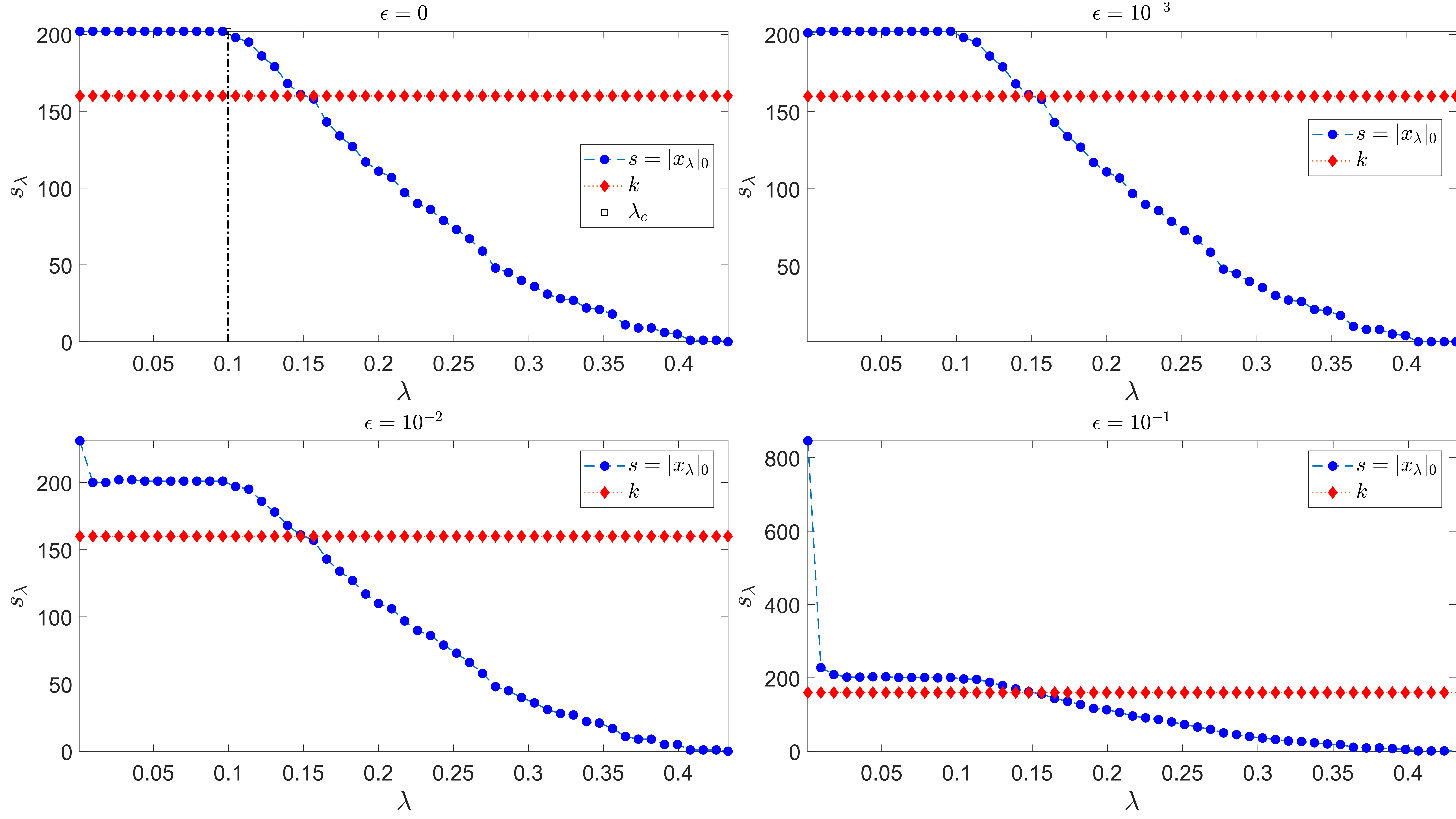}
  \caption{{\small The support for computed minimizer $\slam=\norm{\ulm{\lambda}}{0}$ of $k$-sparse data, $k=160$, peaks at the threshold value of $k_{\textnormal{max}} \sim 215$ when $\lambda$ reaches $\lambda_c\sim 0.11$. This should be compared with the rough upper bound $(1+\deltap)\betaplustheta^2k \leq \vchisquared k$ corresponding to the RIC $\deltap=\vdelta$, and the more realistic bound $\vchisquaredb k$ corresponding to $\delta=\vdeltab$. Observe  (lower figures) that  for exceedingly small $\lambda \ll \epsilon$, there is an additional   growth of order $\displaystyle ~\frac{\epsilon}{\lambda}$.}}
  \label{fig:simull0}
\end{figure}

\begin{figure}[H]
  \centering
  \includegraphics[width=14.8cm,height=12.8cm]{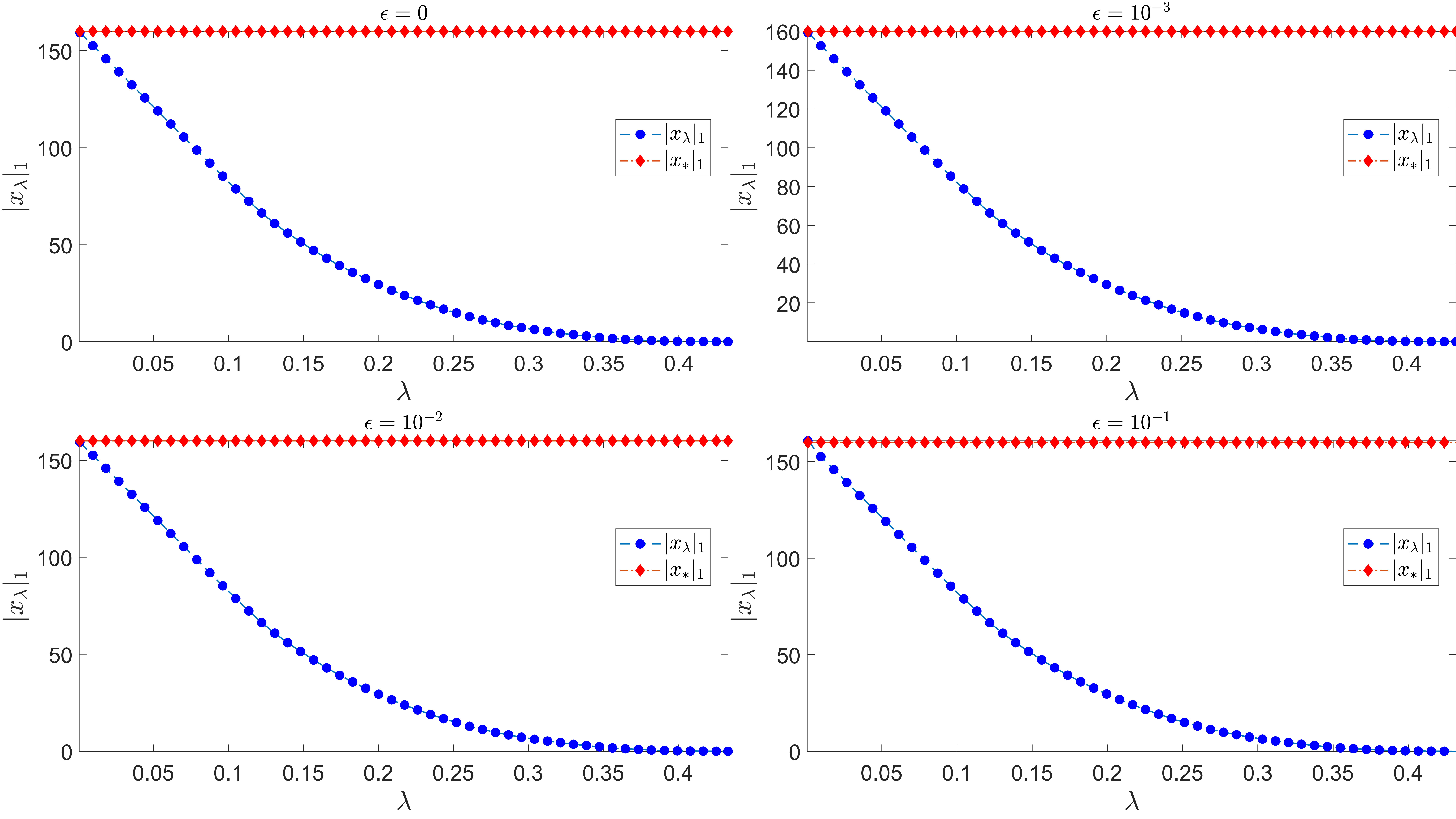}
  \caption{{\small $\ell_1$ norm of $\ulm{\lambda}$ approaches its upper-bound $\norm{\vtarget(\ks)}{1}$ as $\lambda$ decreases.} }
  \label{fig:simull1norm}
\end{figure}

\begin{figure}[h]
  \centering
  \includegraphics[width=14.8cm,height=12.8cm]{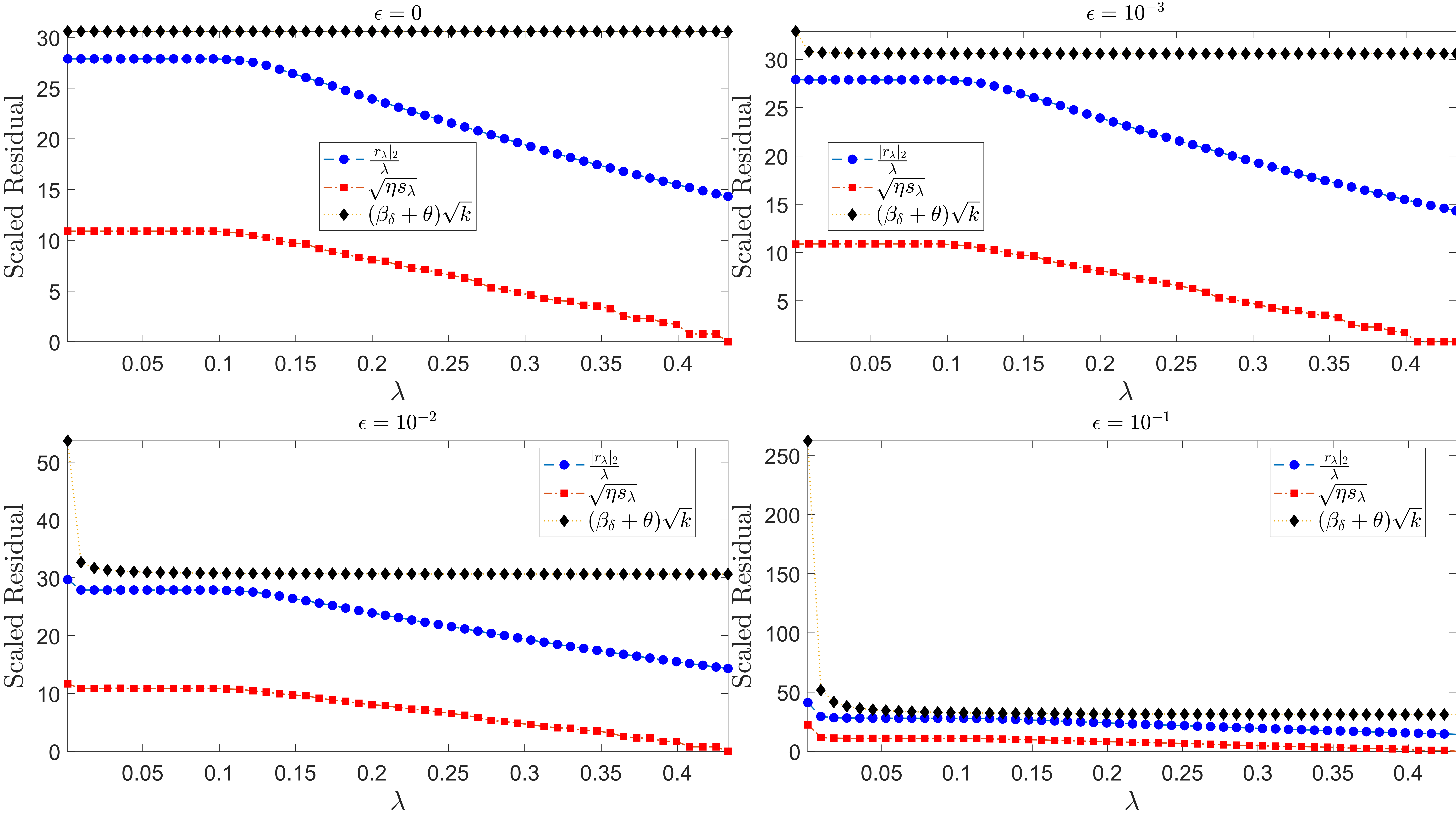}
  \caption{{\small Re-scaled residual $\frac{\norm{\resid{\lambda}}{2}}{\lambda}$ captured between its lower- and upper-bounds  \eqref{eq:lower} and \eqref{eq:res2error}, $\ds \sqrt{\eta\slam}\leq \frac{\norm{\resid{\lambda}}{2}}{\lambda}\leq \betadelta\sqrt{k}+\frac{2\epsilon}{\lambda}\approx \vbetasqrtk$ with $(\eta,\beta,\theta)=(\veta,\vbeta,\vtheta)$ corresponding to $\delta=\vdelta$. It  peaks at a threshold value of ~27, independent of the level of noise. When $\lambda \ll\nl$, there is an additional large term of order $\displaystyle ~\frac{2\epsilon}{\lambda}$.} }
  \label{fig:simulTR}
\end{figure}

\begin{figure}[H]
  \centering
  \includegraphics[width=14.8cm,height=12.8cm]{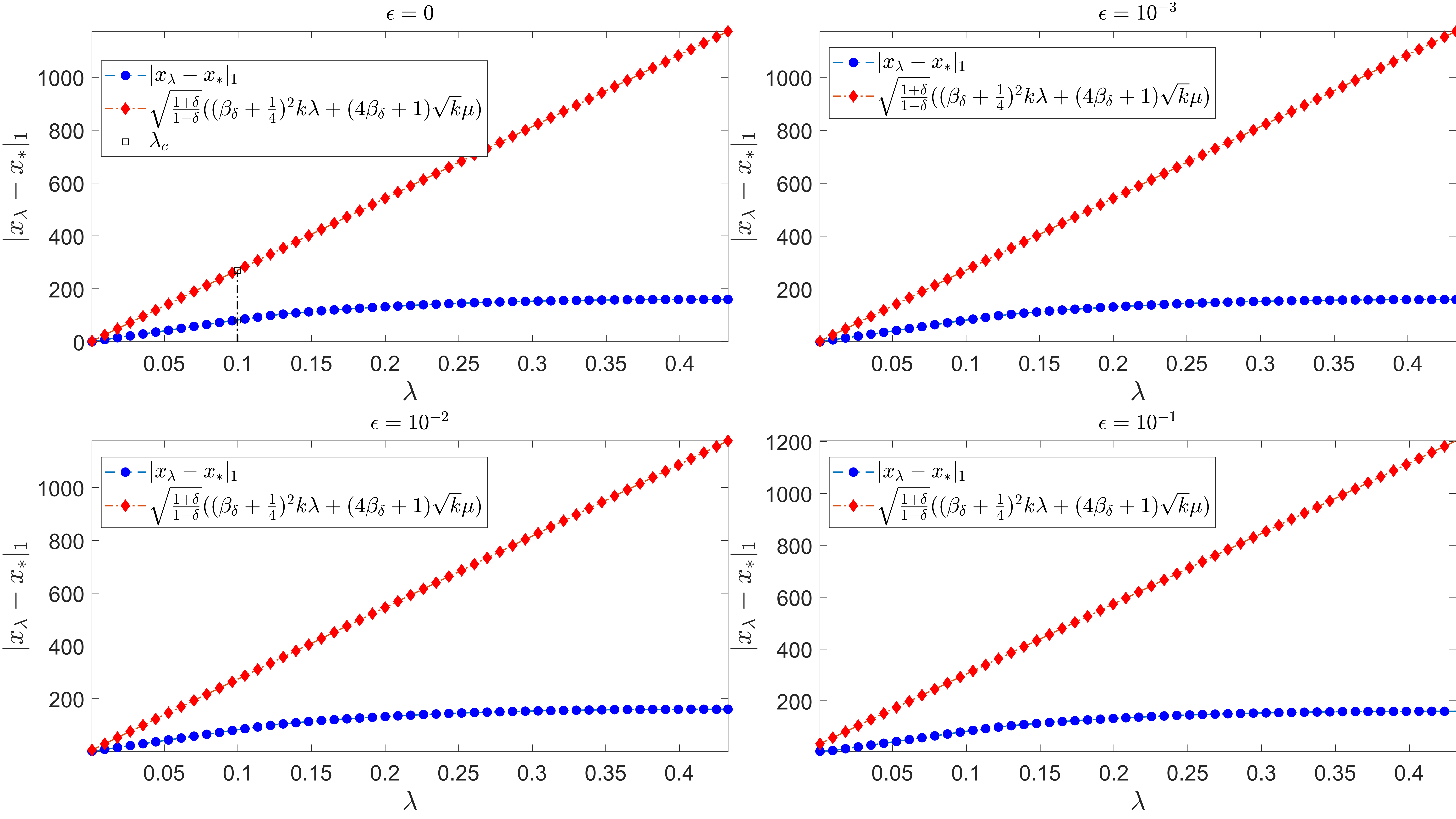}
  \caption{{\small $\ell_1$-error  for recovery of sparse data compared with the upper-bound \eqref{eq:l1boundk=160}.}}
  \label{fig:simull1}
\end{figure}

We report here on our simulations of the unconstrained LASSO \eqref{eq:cs_uncon_l1}, applied  to  the recovery of $k$-sparse data, $\sk=0$, that is $\mu=\epsilon$, with $(k,m,N)=(160,1024,4096)$. 
We consider different \simon{levels} of noise $\epsilon=10^{-3},10^{-2},10^{-1}$, in the corresponding parametric regime \eqref{eq:mu},  $\lambda > \nicefrac{2\mu}{\sqrt{k}}= 0.16\epsilon$.
The results are obtained by averaging 100 observations using  \simon{randomly generated} RNSP${}_{\rho,\tau}$ matrices based on Gaussian distributions. A simple proof of the RIP for such matrices can be found in \cite{baraniuk2008simple}.
The results are  compared with the sparsity bound  of theorem \ref{thm:main}. We note that  our sparsity bound depends in an essential manner on the  the RICs, $1\pm \delta$, in  \eqref{eq:RIP}. 
  The parametric regime in \eqref{eq:whatisdelta}  provides only a rough estimate on the range of allowable RICs, and in particular, does not cover the  parameters used in the simulations below, \cite[\S9.4]{FR2013}.
A detailed study which traces the sharp RICs can be found in  \cite{bah2010improved,bah2014bounds}, but is  beyond the scope of our work. 
We compare the simulations with our sparsity bound based on the RIC $\delta=\vdelta$. This is partly motivated by the result of  \cite{cai2013sparse}
in  which the authors prove an exact BP recovery of $k$-sparse data \simon{from the RIP with} $\delta_{tk} <\sqrt{\nicefrac{(t-1)}{t}}$. In our case, the  computation reported in figure \ref{fig:simull0} indicates  the \emph{actual} sparsity  $\slam<tk$ with $t=2$ which is consistent with  $\delta<\sqrt{\nicefrac{1}{2}}\approx 0.7$. Although the RIC $\delta=\vdelta$ does not provide a tight bound, $\slam<\vchisquared k$, it suffices to detect the correct \emph{behavior} of the LASSO minimizer, reported in figures \ref{fig:simull0}--\ref{fig:simulTR} and \ref{fig:simulEnt}--\ref{fig:simull1}.\newline
  We record here the corresponding parameters involved in our bounds:
  \[
\begin{split}
\betadelta_{{}_{|\delta=\vdelta}}&=\frac{\sqrt{1+\delta}}{\sqrt{1-\delta^2}-\nicefrac{\delta}{4}}
=\vbeta, \quad \rchi_{{}_{|\delta=\vdelta}}=\sqrt{1+\delta}\betaplustheta =\vchi, \quad \eta_{{}_{|\delta=\vdelta}}=\frac{1}{{1+\delta}}=0.59
\end{split}
\]

Our main result on the sparsity of the \LASSO minimizer in theorem \ref{thm:main}
 provides a reasonably accurate information about  the  {behavior} of 
the unconstrained \LASSO minimization \eqref{eq:cs_uncon_l1}.
Figure \ref{fig:simull0} shows the behavior of the support, $\slam=\norm{\ulm{\lambda}}{0}$, starting with $\slam=0$ for $\lambda>\lambda_\infty$ and monotonically increasing  as $\lambda$ decreases all the way to a critical value, $\lambda_c\sim 0.11$, at which point  $s_{\lambda_c}$ reaches its maximal value of $215$. 
This should be compared with our bound $\slam \leq (1+\deltap)\betaplustheta^2k$. For $\deltap=\vdelta$ we have  $ \slam \leq\vchisquared k$, which is  a rough sparsity bound, relative to the actual $\slam\sim 215$. A smaller RIC $\deltap\sim 0.2$ yields a tighter sparsity bound $1.66 k\sim 313$.

\medskip\noindent
Observe that according to  \eqref{eq:xxl}, the $\ell_1$-size of the LASSO minimizer $\ulm{\lambda}$
remains smaller than the target $\norm{\vtarget(\ks)}{1}$, Indeed, as long as the residual $\norm{\resid{\lambda}}{2}>\mu$, then 
\begin{equation}\label{eq:xyz}
\norm{\vtarget(\ks)}{1}  - \norm{\ulm{\lambda}}{1}
   \geq (\norm{\resid{\lambda}}{2}-\mu)\frac{\norm{\resid{\lambda}}{2}}{\lambda}.
   \end{equation}
   This is depicted in figure \ref{fig:simull1norm}: as $\lambda$ decreases, the ratio $\ds \frac{\norm{\resid{\lambda}}{2}}{\lambda}\gtrsim \sqrt{\slam}$ is increasing until $ \norm{\ulm{\lambda}}{1}$ reaches its upper bound of $\norm{\vtarget(\ks)}{1}$.

 \medskip\noindent
Figure \ref{fig:simulTR} shows the \simon{aptitude of the} lower- and upper-bounds of the re-scaled residual  \simon{\eqref{eq:sparsity}}, in capturing the re-scaled residual $\displaystyle \frac{\norm{\resid{\lambda}}{2}}{\lambda}$.  Again, the three quantities increase with deceasing $\lambda$, until $\lambda$ reaches the threshold $\lambda_c$ at which point  the re-scaled residual, $\displaystyle \frac{\norm{\resid{\lambda}}{2}}{\lambda}$, peaks at its maximal value $\sim 27$, in agreement with the  upper-bound \eqref{eq:res2error}\simon{,} 
$\ds 
\frac{\norm{\resid{\lambda}}{2}}{\lambda} < \betadelta\sqrt{k} + 
\frac{2\nl}{\lambda}<\vbetasqrtk + \frac{2\nl}{\lambda}$.

\section{Error bounds}\label{sec:error}
\subsection{$\ell_2$-error bounds}\label{sec:elltwo}
 The sparsity bound \eqref{eq:sparsity} was derived based on a two-sided 
 $\ell_2$-bound of the scaled residual. The latter can be converted into a two-sided $\ell_2$ error bound of $\norm{\ulm{\lambda}-\vtarget(\ks)}{2}$.  
Note that since 
 $\displaystyle 
 \big|\norm{\resid{\lambda}}{2}-\norm{A(\vtarget(\ks)-\ulm{\lambda})}{2}\big| \leq \mu$, 
 then the upper-bound  on $\norm{\resid{\lambda}}{2}$, see \eqref{eq:halftheta}, also bounds the `observed error' $A(\ulm{\lambda}-\vtarget(\ks))$,  
\begin{equation}\label{eq:elltwobound}
\norm{A(\ulm{\lambda}-\vtarget(\ks))}{2} \leq \betaplustheta\sqrt{\ks}\lambda + \mu
\leq \betadelta\sqrt{k}\lambda+3\mu.
\end{equation}
The sparsity of $\ulm{\lambda}$ does not exceed $\klam\leq ([\rchi^2]+1)\ks$ hence  $\ulm{\lambda}-\vtarget(\ks)$ has sparsity $([\rchi^2]+2)\ks$, and the RIP \eqref{eq:RIP}  implies
the $\ell_2$-error upper-bound 
\begin{equation}\label{eq:l2error}
\norm{\ulm{\lambda}-\vtarget(\ks)}{2} \leq \frac{1}{\sqrt{1-\deltap}}(\betadeltap\sqrt{k}\lambda+3\mu), \qquad 
\deltap=\delta_{([\rchi^2]+2)k}.
\end{equation}
In particular, \eqref{eq:l2error} with $\frac{1}{\sqrt{1-\deltap}}\leq \vfracdelta$ and $\betadeltap\leq \vbeta$  corresponding to $\deltap=\vdelta$ yields 
\begin{equation}\label{eq:l2errormax}
\norm{\ulm{\lambda}-\vtarget(\ks)}{2} \lesssim \vfracdeltabeta\sqrt{k}\lambda +\vthreefracdelta\mu.
\end{equation}
This recovers a quantitative version of the  $\ell_2$ upper bound proved under additional condition of an incoherence design assumption in \cite[Theorem 1]{meinshausen2009lasso},  an $\ell_1$-CMSV assumption\footnote{In fact, we slightly improve the quadratic dependence of the bound in \cite[(23)]{tang2011performance} on the $\ell_1$-CMSV constant $\sim \rho^{-2}_{4\ks}$, mentioned in \eqref{eq:CMSV} below.} \cite{tang2011performance}, or restricted eigenvalue bound in \cite[Theorem 11.1]{hastie2015statistical}.\newline
 The upper-bound \eqref{eq:l2error} is sharp in the sense of having a tight    $\ell_2$-\emph{lower bound}:
 since the error $\ulm{\lambda}-\vtarget(\ks)$ is at most  $([\rchi^2]+2)k$-sparse,  we can use the RIP to translate the lower bound \eqref{eq:lower} into an $\ell_2$ lower-bound, 
 \[
 \norm{\ulm{\lambda}-\vtarget(\ks)}{2} \geq \frac{1}{\sqrt{1+\deltap}}\norm{A(\ulm{\lambda}-\vtarget(\ks))}{2} \geq \frac{1}{\sqrt{1+\deltap}}\left(\norm{\resid{\lambda}}{2}-\mu\right) \geq \frac{\sqrt{\klam}\lambda}{1+\deltap}  -\frac{\mu}{\sqrt{1+\deltap}}.
 \]
 We summarize these bounds in the following form.
 \begin{theorem}[{\bf $\ell_2$-bound}]\label{thm:ell2bound}
Fix $\lambda < \lambda_\infty:=|\trans{A}\nresult|_\infty$ and let  $\ulm{\lambda}$ be the  $\slam$-sparse minimizer of the corresponding \LASSO \eqref{eq:cs_uncon_l1}, observed with RIP matrix $A$.
Then
\begin{equation}\label{eq:ell2corbound}
\frac{1}{\sqrt{1+\delta}}\Big(\frac{\sqrt{\klam}\lambda}{\sqrt{1+\delta}} -{\mu}\Big)\leq  \norm{\ulm{\lambda}-\vtarget(\ks)}{2} \leq \frac{1}{\sqrt{1-\deltap}}(\betadelta\sqrt{k}\lambda+ 3\mu), \qquad \deltap=\delta_{([\rchi^2]+2)k}.
\end{equation}
\end{theorem}
  \begin{remark}[{\bf Compared with the $\ell_1$-entropy bound}] 
 The extremal relation  $\innerprod{A_\mS\ulm{\lambda,\mS}}{\resid{\lambda}} = \lambda\norm{\ulm{\lambda,\mS}}{1}$ and the  RIP \eqref{eq:RIP} yield
$\displaystyle 
 \lambda\norm{\ulm{\lambda,S}}{1}\leq \sqrt{1+\delta}\norm{\ulm{\lambda,S}}{2}\norm{\resid{\lambda}}{2}$,
 and hence we end up with a lower-bound involving the $\ell_1$-entropy of $\{\ulm{\lambda,S}\}$,
 \begin{equation}\label{eq:Ent}
 \frac{\norm{\resid{\lambda}}{2}^2}{\lambda^2} \geq\frac{1}{1+\delta}\text{Ent}(\ulm{\lambda,S}) \qquad \text{Ent}(\bx):= \frac{\norm{\bx}{1}^2}{\norm{\bx}{2}^2}.
 \end{equation}
 This bound is tied to  a Null Entropy Property  of $A$ \cite[\S3.2]{andersson2014theorem} or the  $\ell_1$-CMSV constant  $\rho_s(A)$ introduced in \cite{tang2011performance}\footnote{Which is not to be confused with the RNSP parameter in \eqref{eq:RNSP}}
 \begin{equation}\label{eq:CMSV}
 \frac{\norm{\resid{\lambda}}{2}^2}{\lambda^2} \geq \frac{\text{Ent}(\ulm{\lambda,S})}{\rho_s(A)}, \qquad \rho_s(A):=\min_{|\bx|_2=1}\Big\{ |A\bx|_2\, : \, \text{Ent}(\bx)\leq s\big\}.
 \end{equation}
 \bigskip
 
  \begin{figure}[H]
  \includegraphics[width=14.8cm,height=12.8cm]
  {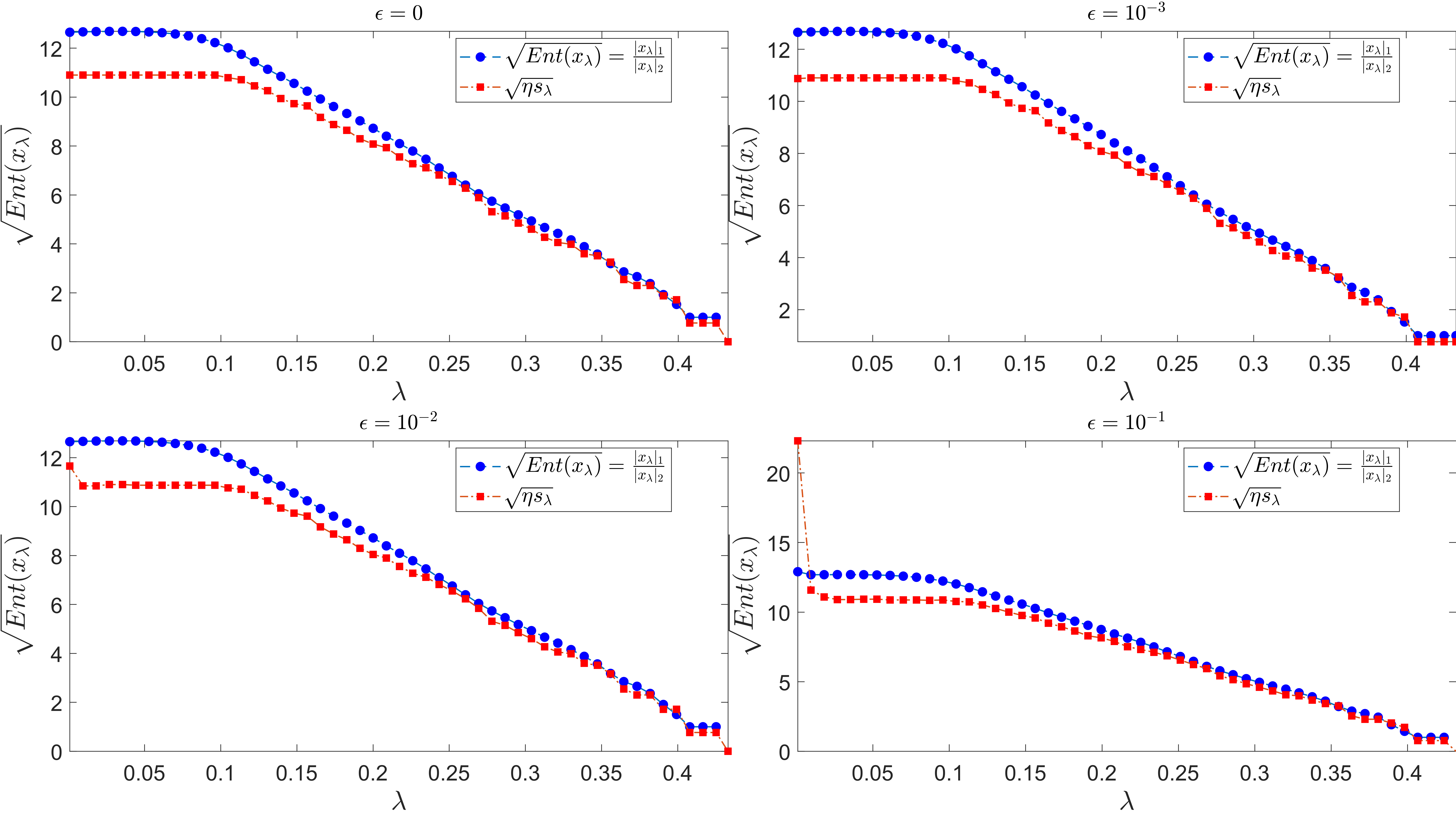}
    \caption{{\small Lower bounds of the re-scaled residual: \eqref{eq:lower} with $\eta:=\frac{1}{1+\delta}=\veta2$ vs. the $\ell_1$-entropy based \eqref{eq:Ent}.}}
  \label{fig:simulEnt}
\end{figure}

\noindent
Clearly, if $\ulm{\lambda}$ has the sparsity $\slam$ then $\text{Ent}(\ulm{\lambda,S})\leq\slam$. Here we note about the reverse implication, namely ---  if the \emph{reverse inequality} holds,  $\text{Ent}(\ulm{\lambda,S})\gtrsim \slam$, then it  would yield our sparsity result of lemma \ref{lem:res3}, based on the lower bound $\displaystyle \frac{\norm{\resid{\lambda}}{2}}{\lambda} \gtrsim \frac{\sqrt{\slam}}{\rho_{\slam}}$. Theorem \ref{thm:main} suggests the lower-entropy bound for the minimizers $\ulm{\lambda}$. Indeed, figure \ref{fig:simulEnt} shows  a remarkable agreement between the lower bound \eqref{eq:lower}  with $\delta=\vdelta$ and  the $\ell_1$-entropy bound \eqref{eq:Ent}, 
 $\text{Ent}(\ulm{\lambda})$, at least before the support of $\ulm{\lambda}$ reaches its peak at $\scr$.  
  \end{remark}

\subsection{$\ell_1$-error bound}\label{sec:ellone}
We recall the  $\ell_2$-bound  \eqref{eq:l2error} which we express in the form
$\ds \norm{\ulm{\lambda}-\vtarget(\ks)}{2}\leq \frac{1}{\sqrt{1-\delta}}(\betadelta+\nicefrac{3}{2}\theta)\sqrt{k}\lambda$. 
Since   $\ulm{\lambda}-\vtarget(\ks)$ has  sparsity of order $\leq k+\rchi^2k$,  we derive the following $\ell_1$-bound.
 \begin{theorem}[{\bf $\ell_1$-error bound}]\label{thm:ell_1$-error estimate}
Fix  $\lambda < \lambda_\infty:=|\trans{A}\nresult|_\infty$ and let $\ulm{\lambda}$ be the  \LASSO minimizer of  \eqref{eq:cs_uncon_l1}, observed with RIP matrix $A$ with  RIC $\delta$ such that \eqref{eq:whatist} holds.  Then  the   following $\ell_1$-error bound holds,
 \begin{equation}\label{eq:l1errormax}
 \begin{split}
\norm{\ulm{\lambda}-\vtarget(\ks)}{1} & \leq \sqrt{\big(1+(1+\delta)\betaplustheta^2\big)k} \norm{\ulm{\lambda}-\vtarget(\ks)}{2}\\
& < \sqrt{1+\delta}(\betadelta+\nicefrac{1}{2}+\theta)\sqrt{k}\frac{1}{\sqrt{1-\delta}}\big(\betadelta+\nicefrac{3}{2}\theta\big)\sqrt{k}\lambda\\
& < \frac{\sqrt{1+\delta}}{\sqrt{1-\delta}}\frac{1}{\lambda}\Big(\big(\betadelta + \nicefrac{1}{2} \big)\sqrt{k}\lambda+ 2\mu\Big)^2.
\end{split}
 \end{equation}
  \end{theorem}
The amplitude of $k\lambda$ in the $\ell_1$-error bound \eqref{eq:l1errormax} is not sharp. For example,   with RIC $\deltap=\delta_{\vchisquared k}<\vdelta$ we have $\betadelta>2$ in which case, omitting the negligibly small $\mu^2/\lambda$ terms,  one ends up with the improved bound 
\begin{equation}\label{eq:l12}
\norm{\ulm{\lambda}-\vtarget(\ks)}{1} < \frac{\sqrt{1+\delta}}{\sqrt{1-\delta}}\Big(\big(\betadelta + \nicefrac{1}{4}\big)^2 k\lambda  +(4\betadelta+1)\sqrt{k}\mu\Big) < \vlonebound k\lambda + \veightbeta\sqrt{k}\mu.
\end{equation}

We conclude with an alternative derivation of an $\ell_1$-error bound. To this end, we recall  the RNSP bound  \cite[Theorem 4.20]{FR2013}, which states that 
for all $\mK \subset \{1,2\,\ldots, N\}$ of size$\leq k$ and for any $\bu$, $\bv \in \realR{N}$, the following holds, 
  \[
    \norm{\bu - \bv}{1} \le \frac{1 + \rho}{1 - \rho}\big(\norm{\bu}{1} - \norm{\bv}{1} + 2\norm{\bv_{\compl{\mK}}}{1}\big) + \frac{2\tau}{1 - \rho}\norm{A(\bu - \bv)}{2}, \qquad |\mK|\leq k.
  \]
Using it   with $(\bu,\bv)=(\ulm{\lambda},\vtarget(\ks))$ and $\mK = \supp{\vtarget(\ks)}$ yields
  \begin{equation}\label{eq:tol1}
 \norm{\ulm{\lambda} - \vtarget(\ks)}{1} \le \frac{1 + \rho}{1 - \rho}\Big(\norm{\ulm{\lambda}}{1} - \norm{\vtarget(\ks)}{1} \Big) + \frac{2\tau}{1 - \rho}\norm{A(\simon{\ulm{\lambda} - \vtarget(\ks)})}{2}.
  \end{equation}
  Now, using \eqref{eq:xxl} to bound the term inside the first parenthesis on the right, and as before, noting that the second term does not exceed $\norm{A(\ulm{\lambda} -\vtarget(\ks))}{2}\leq \norm{\resid{\lambda}}{2}+\mu$,  we find
\[
\norm{\ulm{\lambda} - \vtarget(\ks)}{1} \le \frac{1+\rho}{1 - \rho}\Big\{ \norm{\resid{\lambda}}{2} \Big( \frac{2\tau}{1+\rho} + \frac{\mu}{\lambda} - \frac{\norm{\resid{\lambda}}{2}}{\lambda}\Big) + \frac{2\tau}{1+\rho}\mu\Big\}.
\]
Given  the RNSP parameters \eqref{eq:RNSP},  $\ds {2\tau}=2\beta\sqrt{k}$ and  $\ds \frac{\mu}{\lambda}<\frac{\theta}{1+\rho}\sqrt{k}$,
the last bound yields
\begin{equation}\label{eq:l1error}
\norm{\ulm{\lambda} - \vtarget(\ks)}{1} \leq \frac{1}{1 - \rho}\left\{ \norm{\resid{\lambda}}{2}\left({2\betaplustheta\sqrt{k}}-(1+\rho)\frac{\norm{\resid{\lambda}}{2}}{\lambda}\right)+{\betadelta\theta}\simon{k}\lambda \right\}.
\end{equation}
Viewed as quadratic in $\displaystyle \frac{\norm{\resid{\lambda}}{2}}{\lambda}$,  the first expression   on the right  admits a  maximal value  
 $\displaystyle \frac{\betaplustheta^2 }{1+\rho}k\lambda$, and we finally end up with
 \begin{equation}\label{eq:lamerror}
 \begin{split}
  \norm{\ulm{\lambda} - \vtarget(\ks)}{1}& \leq \frac{1}{1-\rho^2}\left( (\betadelta+{\theta})^2\ks \lambda + (1+\rho)\betadelta\theta k\lambda \right)    \leq \frac{(\betadelta+2\theta)^2}{1-\rho^2}k\lambda.
   \end{split}
 \end{equation}
 This recovers  the $\ell_1$-bound  of order ${\mathcal O}(\ks\lambda)$ as in \eqref{eq:l1errormax}.
However, since the $\ell_1$ bound \eqref{eq:lamerror} involves  the value of  $\nicefrac{1}{(1-\rho)^2}$, it is therefore limited to the RIC $\delta < \nicefrac{4}{\sqrt{41}}$ where  $\rho$ approaches 1.  

\ifx
Yet, the error bound \eqref{eq:l1error} provides a more precise information on the \emph{behavior} of  $\ell_1$-error  in terms of the size of the support $\slam$, which refines  the uniform-in-$\slam$ error bounds in theorem \ref{thm:ell_1$-error estimate} and \eqref{eq:lamerror}.
\begin{theorem}[{\bf $\ell_1$-error bound revisited}]\label{thm:ell1}
Fix $\lambda < \lambda_\infty$ and let $\ulm{\lambda}$ be the \LASSO minimizer \eqref{eq:cs_uncon_l1} with small scale $\mu$ in \eqref{eq:mu} and with observing $A$s satisfying RNSP${}_{\rho,\tau}$ of order $k$. The following $\ell_1$-error bound holds, 
\begin{subequations}\label{eqs:l1generror}
\begin{equation}\label{eq:l1generror}
\norm{\ulm{\lambda} - \vtarget(\ks)}{1} \le
 Q(\slam)k\lambda + \frac{\beta}{1-\rho}\sqrt{k}\mu, \qquad \slam=\norm{\ulm{\lambda}}{0}.
\end{equation}
Here,  $Q(s)$ is a piecewise-quadratic   in $\sqrt{\nicefrac{s}{k}}$, 
\begin{equation}\label{eq:l1imperror}
Q(s):=\left\{\begin{array}{ll}
\displaystyle \frac{\rchi^2}{1-\rho^2}, & \displaystyle \nicefrac{s}{k}< \frac{\rchi^2 }{(1+\rho)^2}, \\ \\ 
\displaystyle C_\rho\sqrt{\nicefrac{s}{k}}\Big(\frac{2\rchi}{1+\rho}  - \sqrt{\nicefrac{s}{k}}\Big), \ \ \ 
& \displaystyle \frac{\rchi^2 }{(1+\rho)^2} \leq \nicefrac{s}{k}\leq\rchi^2,\\ \\ 
\rchi^2 & \displaystyle \rchi^2\leq \nicefrac{s}{k}\leq ([\rchi^2]+1).
\end{array}\right. 
\end{equation}
\end{subequations}
\end{theorem}
\noindent
\begin{proof}
Set $\displaystyle \zeta:=(1+\delta)\frac{\norm{\resid{\lambda}}{2}^2}{\lambda^2}$. Then  the  error bound \eqref{eq:l1error} is expressed as 
\[
\norm{\ulm{\lambda} - \vtarget(\ks)}{1} \leq \frac{C_\rho}{1+\delta} 
\sqrt{\nicefrac{\zeta}{k}}\Big(\frac{2\sqrt{1+\delta}(\beta+\nicefrac{\theta}{2})}{1+\rho}-\sqrt{\nicefrac{\zeta}{k}} \Big)k\lambda+ \frac{\beta \theta}{4(1-\rho)}k\lambda.
  \]
  The quadratic expression on the right, $\ds Q(\zeta)=C_\rho\sqrt{\nicefrac{\zeta}{k}}\Big(\frac{2\rchi}{1+\rho}-\sqrt{\nicefrac{\zeta}{k}} \Big)$,  has the obvious upper-bound $\ds C_\rho\frac{\rchi^2}{(1+\rho)^2}$ for all $\zeta$'s, which is the first part of \eqref{eq:l1imperror}.
On the other hand, when $\ds \slam\geq \frac{\rchi^2 k}{(1+\rho)^2}$ then the lower-bound \eqref{eq:lower} implies
 $\ds \zeta  > \slam \geq\frac{\rchi^2 k}{(1+\rho)^2}$, 
and since $Q(\zeta)$ is  decreasing  for $\zeta$ in that range, $\ds \zeta\geq \frac{\rchi^2 k}{(1+\rho)^2}$, it follows that 
\[
\norm{\ulm{\lambda} - \vtarget(\ks)}{1} <  Q(\zeta)k\lambda +\frac{\beta}{1-\rho}\sqrt{k}\mu \leq  Q(\slam)k\lambda+ \frac{\beta}{1-\rho}\sqrt{k}\mu,
\] 
which proves the second part of \eqref{eq:l1imperror}. Finally, when $\slam \geq \rchi^2k$, approaching its maximal value $([\rchi^2]+1)k$ indicated in \eqref{eq:smax}, we find
\[
Q(\slam)=C_\rho\rchi\left(\frac{2}{1+\rho}\rchi-\rchi\right)= \rchi^2,
\]
proving the third part of \eqref{eq:l1imperror}.
\end{proof}
\fi
\subsection{Numerical simulations}\label{sec:simul}

We report  on the error behavior in our simulations of the unconstrained LASSO \eqref{eq:cs_uncon_l1}, applied  to  the recovery of $k$-sparse data, $\sk=0$, that is $\mu=\epsilon$, with $(k,m,N)=(160,1024,4096)$. The results are obtained by averaging 100 observations using \simon{randomly generated} RNSP${}_{\rho,\tau}$ matrices based on Gaussian distributions. 
We  compare the $\ell_1$-error with the error bound \eqref{eq:l12}
\begin{equation}\label{eq:l1boundk=160}
\norm{\ulm{\lambda}-\vtarget(\ks)}{1} \leq   \vlonebound*160 \lambda + \veightbeta\sqrt{160}\,\nl.
\end{equation}

\begin{figure}[H]
  \centering
  \includegraphics[width=14.8cm,height=12.8cm]{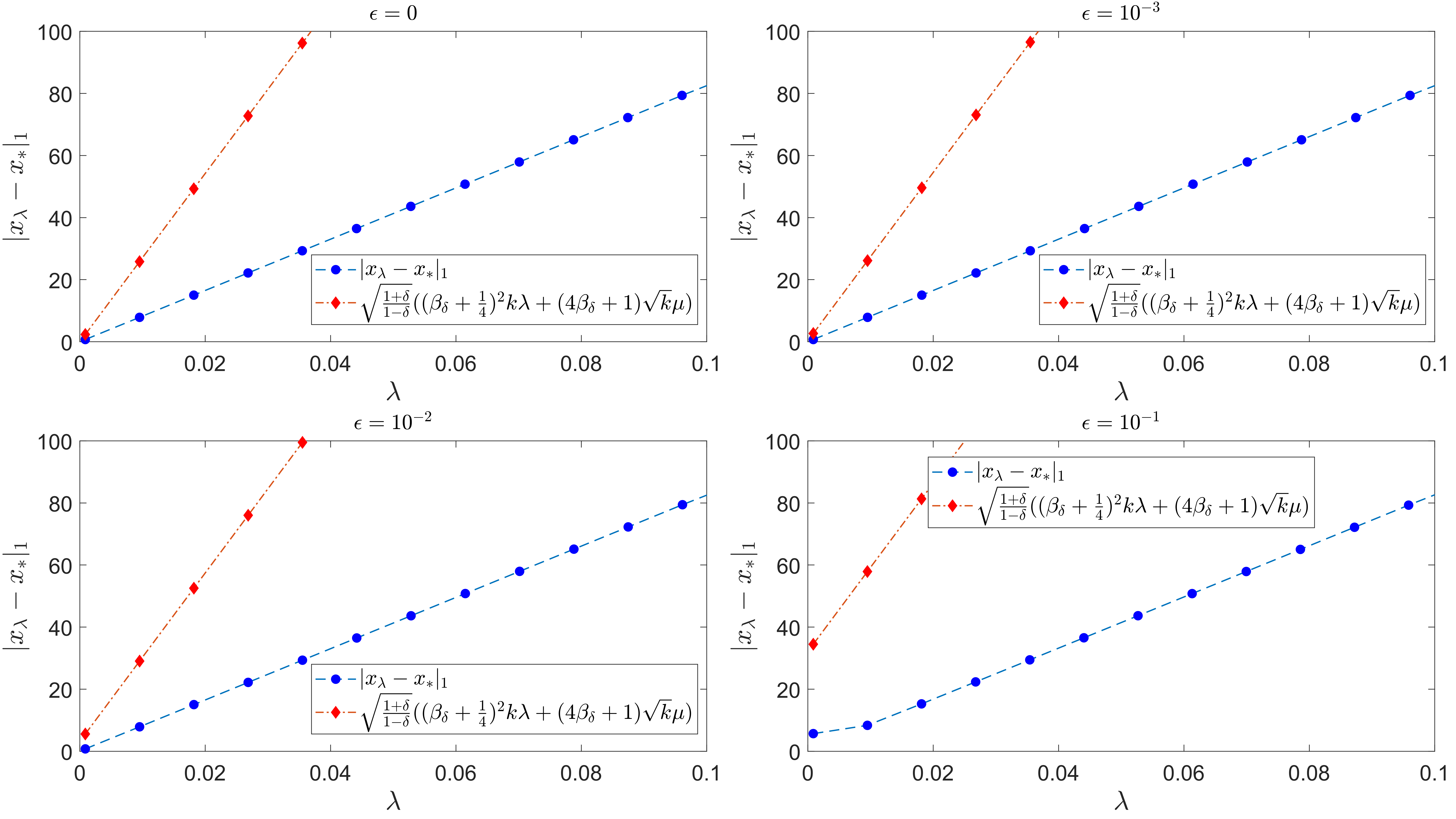}
  \caption{{\small The $\ell_1$-error compared with the  upper-bound \eqref{eq:l1boundk=160} zoomed near $\lambda=0$.}}
  \label{fig:simull1zoomed}
\end{figure}

\medskip\noindent
 {\bf Acknowledgment}. We are indebted to Wolfgang Dahmen for comments which greatly improved the presentation of material of this paper. 

\bibliographystyle{siam}
\bibliography{reference}

\end{document}